\documentclass{aa}  
\graphicspath{{images/}}
\usepackage{graphicx}
\usepackage{txfonts}
\usepackage{hyperref}

\hypersetup{
    colorlinks=true, 
    linkcolor=blue, 
    citecolor=blue,
    urlcolor=blue, 
}

\begin{document}

   \title{Shock imprints on the radio mini halo in RBS~797}

   \author{A. Bonafede\inst{1,2},  M. Gitti\inst{1,2}, N. La Bella\inst{1,3}, N. Biava\inst{1,2}, F. Ubertosi\inst{1,4}, 
   G. Brunetti \inst{2},
   G. Lusetti \inst{5,1}, M. Brienza\inst{4,1}, C. J. Riseley\inst{1,2,8}, C. Stuardi\inst{1,2},  A. Botteon\inst{3}, A. Ignesti\inst{6}, H. R\"ottgering\inst{7}, R. J. van Weeren\inst{7}
          }

   \institute{DIFA - Universit\`a di Bologna, via Gobetti 93/2, I-40129 Bologna, Italy\\
              \email{annalisa.bonafede@unibo.it}
         \and INAF - IRA, Via Gobetti 101, I-40129 Bologna, Italy; 
         \and Department of Astrophysics, Institute for Mathematics, Astrophysics and Particle Physics (IMAPP), Radboud University, P.O. Box 9010, 6500 GL Nijmegen, The Netherlands
         \and INAF - OAS, Via Gobetti 93/2, I-40129 Bologna, Italy.
             \and University of Hamburg, Gojenbergsweg 112, 21029 Hamburg, Germany.
             \and  INAF - Astronomical Observatory of Padova, vicolo dell'Osservatorio 5, IT-35122 Padova, Italy 
             \and
             Leiden Observatory, Leiden University, PO Box 9513, 2300 RA Leiden, The Netherlands.\\
            }
   \date{Received September 15, 1996; accepted March 16, 1997}

  \abstract
  {}
  {In this work, we analysed  new LOw Frequency ARray observations of the mini halo in the cluster RBS~797, together with archival Very Large Array observations and the recent {\it Chandra} results.  This cluster is known to host a powerful active galactic nucleus (AGN) at its centre, with two pairs of jets propagating in orthogonal directions. Recent X-ray observations have detected three pairs of shock fronts within 125 kpc from the cluster centre, connected with the activity of the central AGN. Our aim is to investigate the connection between the mini halo emission and the activity of the central source. }
  {We have used different methods to separate the emission of the central source from the diffuse mini halo emission, and we have derived the radial spectral index trend of the mini halo.}
  {We find that the diffuse radio emission is elongated in different directions at 144 MHz (east-west) with respect to 1.4 GHz (north-south), tracing the orientation of the two pairs of jets.
 The mini halo emission is characterised by an average spectral index $\alpha=-1.02\pm 0.05$. The spectral index profile of the mini halo shows a gradual flattening from the centre to the periphery. Such a trend is unique among the mini halos studied to date, and resembles the
 spectral index trend typical of particles re-accelerated by shocks. However, the estimated contribution to the radio brightness profile coming from shock re-acceleration is found to be insufficient to account for the radial brightness profile of the mini halo. } 
 {We propose three scenarios that could explain the observed trend: (i) the AGN-driven shocks are propagating onto an already existing mini halo, re-energising the electrons and leaving clear imprints in the mini halo spectral properties. We estimate that the polarisation induced by the shocks could be detected at 6 GHz and above; (ii) we could be witnessing turbulent re-acceleration in a high magnetic field cluster; and (iii) the mini halo could have a hadronic origin, in which the particles are injected by the central AGN and the diffusion coefficient depends of the cosmic ray proton momentum.
 A future observation in polarisation would be fundamental to understand the role of shocks and the magnetic field.
 }


\titlerunning{Radio mini halo in RBS~797}
\authorrunning{A. Bonafede et al.}

   \maketitle
%

   
\section{Introduction}

Diffuse non-thermal emission has been observed in a number of galaxy clusters revealing the presence of relativistic electrons and magnetic fields in the intra-cluster medium  (ICM) \cite[see, e.g. ][for a recent rerview]{vanWeeren19}.  Mini halos are synchrotron radio sources found at the centre of cool-core galaxy clusters and surrounding the central radio-loud active galactic nucleus (AGN). Mini halos have typical sizes of a few hundred kiloparsecs, a low surface brightness at gigahertz frequencies \citep[e.g.][]{Giacintucci14}, and a steep spectrum:\footnote{In this work, we use the convention $ S ( \nu ) \propto \nu^{\alpha}$} $\alpha < -1$. 
As the radiative lifetime of the electrons injected in the ICM by the AGN jets is shorter than the time required to reach the mini halo external regions, some (re-)acceleration or an in situ production  process must be in place.\\
Two main models have been proposed to explain mini halos: hadronic models \citep{pe04} and re-acceleration models \citep{Gitti2002,ZuHone2013}.
In the hadronic model, the electrons responsible for the synchrotron emission would originate as a product of the decay from cosmic ray proton-thermal proton collisions. In the re-acceleration scenario, cosmic ray electrons would be continuously re-accelerated by turbulence, injected in the ICM by either AGN feedback and/or the sloshing of the cluster core after a minor merger. Recent works indicate that
none of the two models alone, in their simplest version, can easily explain all of the observed features \citep[e.g.][]{Riseley22,Biava21b}. 

Cold fronts, detected in an increasing number of clusters with a mini halo, support the idea that turbulence is induced by a gas sloshing motion \citep{Mazzotta2008,ZuHone2013,Giacintucci2014}. Other studies indicate that the central AGN may be the source of turbulence \citep{Bravi2016}. 
The connection between the central AGN and the mini halo emission has been investigated by \cite{Gendron17}, who analysed the mini halo in the Perseus galaxy cluster. \cite{Gendron17} found that both the AGN activity and the sloshing motion of the cluster core influence the mini halo emission. More recently, \cite{Richard-Laferriere2020} have found two correlations that suggest a close link between the AGN activity and the mini halo: (i) a correlation between the power of the mini halo and the power of the AGN steep-spectrum component, and (ii) a correlation between the power of the mini halo and the power of the inner X-ray cavities of each system.
In addition, \citet{Seth22} found that also non-central sources may have a significant impact, balancing ICM cooling in clusters.

RBS~797 is a  massive cool-core cluster, which  is known to host a radio mini halo \citep{Gitti2006} and a powerful AGN at its centre.
In Tab. \ref{tab:cluster} we list the main cluster properties.
 RBS~797 was first discovered with the ROSAT All-Sky Survey \citep{RASS}. \textit{Chandra} observations revealed the presence of two X-ray cavities \citep{Schindler2001} located at the north-east and south-west (east-west cavities) with respect to the central brightest cluster galaxy (BCG). The central BCG is a powerful AGN, which has been studied in the radio band by  \citet{Gitti2006,2011cavagnolo,Doria2012,Gitti13}. These studies have detected two jets from the central AGN, oriented in the east-west direction. The lobes associated with these jets are co-spatial with the cavities detected in the X-rays. A second pair of jets has been detected, departing from the AGN and oriented north-south, that is, orthogonal with respect to the east-west lobes. 
 Recently, deep {\it Chandra } observations have revealed a second pair of cavities in the north-south direction, co-spatial with the north-south lobes of the AGN. Interestingly, both cavity pairs are at the same distance from the cluster centre (i.e. $\sim27$ kpc,  \citealt{Ubertosi21}), making RBS797 the only system so far with four equidistant radio-filled X-ray cavities. The authors could not discriminate between coeval outbursts (north-south and east-west on a timescale $\Delta t < 10$ Myr) or a rapid re-orientation of the jets. The same deep {\it Chandra } observations have led to the discovery of three pairs of shock fronts, found at projected distances of 50 kpc, 80 kpc, and 130 kpc from the cluster centre \citep{Ubertosi22}.\\
It has been proven that sloshing motions can produce multiple shocks \citep{Fujita2004}. However, there are no detected signs of sloshing in the ICM and the detection of X-ray cavities suggests that the AGN activity is likely responsible for the shocks. Hence, we follow in this work the interpretation of \citet{Ubertosi22}, and assume that shocks are due to the AGN activity.

A radio mini halo has been found in RBS~797 at 1.425 GHz surrounding the central AGN \citep{Gitti2006}, with a largest lines scale of $\sim 200$ kpc, that is, comparable with the  full length of the cooling region. 
\cite{Ubertosi22} found that the mini halo emission at 1.425 GHz is confined within the middle pair of shocks. 
\citet{Ignesti20} have analysed the correlation between the thermal and non-thermal emission of the cluster. They found a super-linear scaling of the radio brightness $I_R$ with respect to the X-ray brightness $I_X$: $I_R \propto I_X^{1.27\pm 0.12}$, which was obtained excluding the central AGN. \citet{Ignesti20} conclude that the mini halo emission is consistent with a hadronic origin, provided that the central cluster magnetic field, $B_0,$ is $B_0 \sim 10 \, \mu$G for a magnetic field whose energy scales as the cluster thermal energy.
\\
In this work, we present new LOw Frequency ARray  High Band Antennas \citep[LOFAR HBA;][]{vanHaarlem2013} observations of the cluster RBS~797, and we use archival Very Large Array (VLA) data to investigate the link between the mini halo properties and the AGN activity. Thanks to its sensitivity and working at low radio frequencies, LOFAR observations are ideal to derive spectral index information of the emission detected at higher frequencies by the VLA.\\
The paper is organised  as follows:
  in Section \ref{sec:obs}, we present the observations used in this work. Results from the new LOFAR observations are shown in Sec. \ref{sec:res}. In Sec. \ref{sec:fits} we show the different methods we have used to separate the emission of the mini halo from the AGN emission, and we analyse the radio multifrequency emission of the mini halo in Sec. \ref{sec:spix}. We discuss our results in Sec. \ref{sec:discussion} and conclude in Sec. \ref{sec:concl}. 
RBS~797 is  at  a  redshift  of 0.354. We assume a cosmology with $ {\rm{H_{0}}} = 70 \rm {\, km \ s^{-1} Mpc ^{-1}}, \rm{\Omega_{m}} = 0.3 $, and $ \rm{\Omega_{\Lambda}} = 0.7$, for which 1\arcsec corresponds to 4.98 kpc.

\begin{table}[]
    \centering
     \caption{Cluster RBS~797}
     \renewcommand\arraystretch{1.2}
    \begin{tabular}{ c c c c}
    \hline
    \hline
     RA     & DEC      & z   & $M_{500}$  \\
    J2000 & J2000  &     &  $10^{14} \rm{M_{\odot}}$  \\
   \hline
     09h47m13.0s & +76d23m14s  & 0.354 & 5.6$\pm$0.5 \\
  \hline
  \hline
  \multicolumn{4}{l}{\scriptsize Col. 1 and 2: Cluster's right ascension and declination;}\\
  \multicolumn{4}{l}{\scriptsize Col. 3 cluster redshift \citep{Sanders11};}\\
  \multicolumn{4}{l}{\scriptsize Col 4: cluster mass within $R_{500}$ \citep{Planck16}}\\
    \end{tabular}
   
    \label{tab:cluster}
    
\end{table}

\begin{figure*}
\centering
\includegraphics[width=0.7\textwidth]{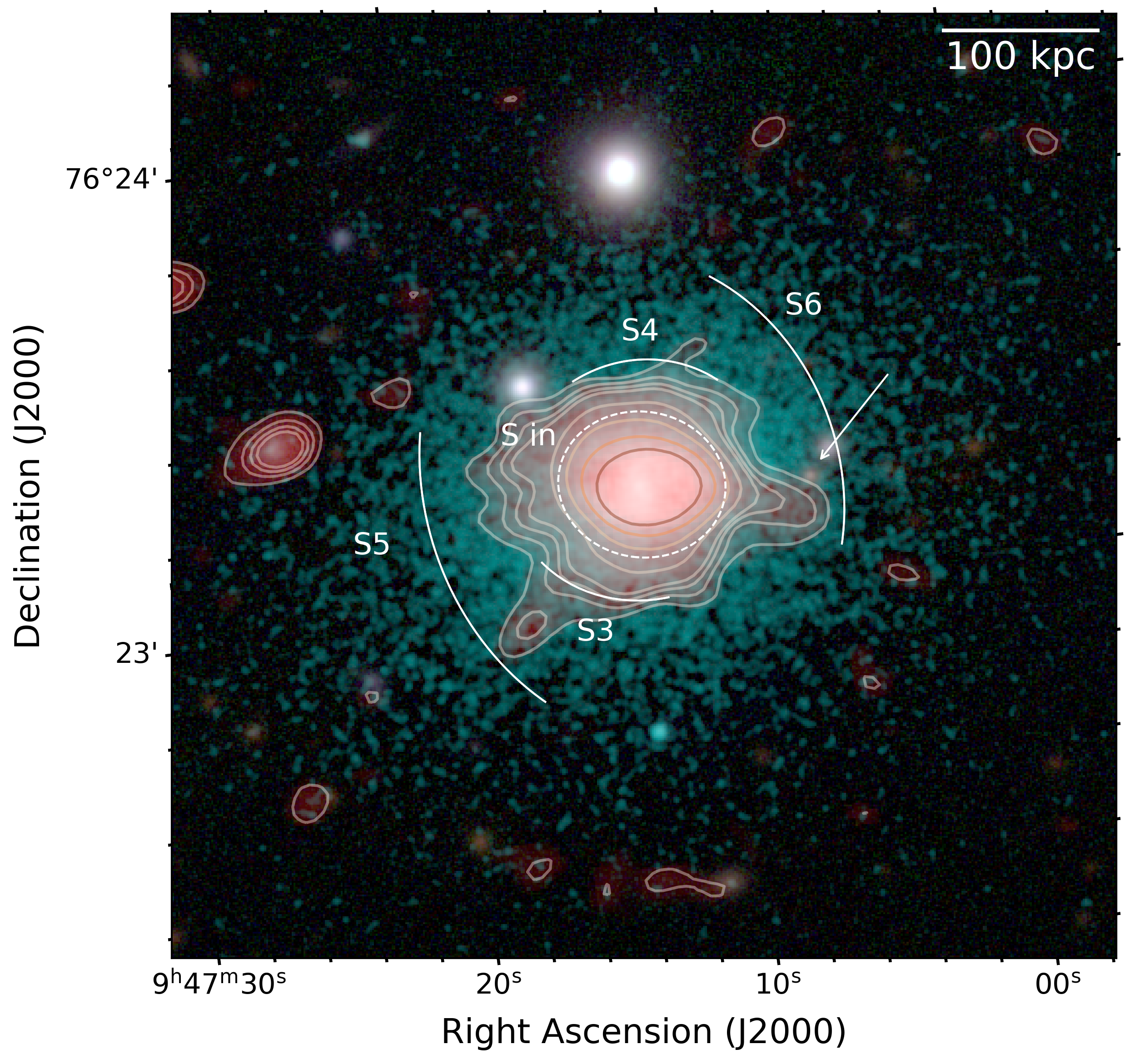}
\caption{Composite optical (Pan-STARRS), X-ray (\textit{Chandra}), and radio (LOFAR) image.
The LOFAR HBA image of the cluster RBS~797 is shown in red and in contours. The noise is $\sigma_{rms}= 0.1~\rm{mJy~beam^{-1}}$, and the beam is 7.0\arcsec$\times$4.7\arcsec. Contours start at 3$\sigma_{rms}$ and increase by a factor 2 each.
The Chandra image is displayed in cyan colours, the Pan-STARRS image is a composite g,r,i image. 
The arcs labelled S3-S6 and the dashed ellipse labelled S$_{\rm{in}}$ refer to the shock fronts detected by \cite{Ubertosi22}. 
The white arrow points the position of the plume (see Sec. \ref{sec:res}).}
\label{fig:LOFAR_all}
\end{figure*}

\section{Observations and data reduction}
\label{sec:obs}
\subsection{LOFAR observations}
The cluster RBS~797 has been observed by the LOw Frequency ARray (LOFAR, \citealt{vanHaarlem2013}) using the same observational setup and data calibration scheme as done in the LOFAR Two Meter Sky Survey (LoTSS) Data Release 2 \citep{Shimwell2017,Shimwell2019,Shimwell22,Tasse21}. In the following, we outline the main steps of the observations and data processing, and refer the reader to  \citet{Shimwell22,Tasse21} for further technical details. We also note that the same observations are analysed in \citet{Biava_sub} to search for extended emission outside the cluster core.\\
RBS~797 has been observed with LOFAR using the HBA antennas in the DUAL\_INNER mode configuration. Observations are centred at 144\,MHz and have a 48\,MHz total bandwidth. The target RBS~797 has been observed for 8h, book-ended by 10 min observations of a calibrator (3C295 and 3C147). Data have been recorded with a time and frequency resolution of 1s and 3.051 kHz, respectively. After initial flagging, data have been averaged by a factor of 4 in frequency.\\
We have corrected for the ionospheric Faraday rotation, clock offsets, instrumental XX and YY phase offsets, and time-independent amplitude solutions using the calibrators gains and applying them to the target.
The \cite{ScaifeHeald12} flux density scale is used for the amplitude calibration, although amplitude gains are refined during the self-calibration process.
\\
The directional-independent calibration of the target data was performed using the $\mathtt{Prefactor}$ pipeline \citep{DeGasperin2019, VanWeeren2016}.  
We have used the TGSS-ADR1 to obtain an initial model of the sky and perform a first phase-calibration of the target field. 
After this initial direction-independent calibration, data have been averaged into 24 visibility files, each having a bandwidth of 1.953\,MHz with a frequency resolution of 97.6\,kHz and a time resolution of 8s. 
To correct for differential amplitude and phase effects, we have used the direction-dependent calibration pipeline used by LoTSS, that includes the usage of $\mathtt{KillMS}$ \citep{Smirnov15} for direction-dependent gains and $\mathtt{DDFacet}$ \citep{Tasse18} to apply the direction-dependent solutions during imaging. Direction-dependent solutions have been obtained and applied for 45 directions within the field of view. \\
To further improve the quality of the target images and to speed-up the imaging process, we have applied the so called ``extraction and self-calibration" method, as already done and described by \citet{vanWeeren21,Botteon22,Biava_sub}, . Direction-dependent solutions are used to subtract the sky emission outside a square region of 20\arcmin ~from the cluster centre from the UV visibilities.  The extracted dataset is corrected for the LOFAR station beam and self-calibrated in amplitude and phase (specifically, 4 phase and 6 phase and amplitude loops). $\mathtt{WsClean}$ \citep{wsclean} is used for imaging and $\mathtt{DPPP}$ \citep{DPPP} is used to obtain the calibration solutions of the extracted dataset.\\
We have imaged the cluster with different parameters to gain sensitivity towards compact and diffuse emission. In Tab. \ref{tab:sources}, we list the UV-range, Briggs Robust weighting scheme \citep{Briggs}, and UV-taper that we have applied to the final self-calibrated dataset.\\
In order to correct for possible offset in the absolute flux-scale, we have followed the approach used by the LOFAR survey team \citep{Shimwell22}, deriving a correction factor of 1.46, that we have applied to the images. We assume a 10\% uncertainty of the absolute flux scale at LOFAR frequencies \citep{Shimwell22}.

\subsection{VLA observations}
To analyse the spectral properties of the diffuse emission, we have used the radio observations in L band published by \cite{Gitti2006} and \cite{Doria2012}. We refer the reader to the original paper for details about the observations and calibration. We have combined data in the A, B, and  C configurations, as done by \citet{Doria2012} and we have imaged them using $\mathtt{WsClean v3.1.0}$  \citep{wsclean}. We assume a 5\% uncertainty on the absolute flux scale at 1.4 GHz \citep{PerleyButler17}.

\begin{table*}
\caption{Image details}
\renewcommand\arraystretch{1.2}
\begin{tabular}{cccccccc}
\hline\hline
 Freq &UV-range& Robust & UV-taper &$\theta_{FWHM}$ & $\sigma_{rms}$ & Fig.  & Notes\\
MHz & $\lambda$ &     &           & $'' \times ''$ & mJy/beam  & &  \\
\hline
144 & $>$80 & -0.5 & - & 7.0$\times$4.7 & 0.1 & \ref{fig:LOFAR_all} & \\
1425 & - & -0.5 & - & 5.1$\times$4.4 & 0.02 &  not shown   & \\
144  & $> 8 \cdot 10^3$   & -2  &  -  & 3.2 $\times$ 2.2 & 0.4 &  \ref{fig:VLA_LOFAR_HL}, bottom panel (cyan contours) &   \\
1425    & $>10^4$   & -2  &  -  &  2.4$\times$1.2   & 0.02 & \ref{fig:VLA_LOFAR_HL}, top panel(cyan contours) &\\
144  & $>$160   & -0.5  &  -  & 7 $\times$ 5  & 0.1 &   \ref{fig:VLA_LOFAR_HL}, bottom panel  & UV-subtracted\\
1425 & $>$160   & -0.5  &  3  & 7 $\times$ 5  & 0.02 & \ref{fig:VLA_LOFAR_HL},top panel & UV-subtracted \\

\hline\hline
\multicolumn {8}{l}{\scriptsize Col 1: Image frequency; Col 2: Image UV-range ; Col 3: Briggs Robust parameter;Col 4: UV-taper used in the image;  Col 5: Restoring beam; Col 6: rms noise; Col 7: Figure; }\\
\multicolumn{8}{l}{\scriptsize Col 8: additional notes.}
\end{tabular}\\ 
\label{tab:images}
\end{table*}

\begin{table*}
\centering
\caption{Sources' properties}
\renewcommand\arraystretch{1.2}
\begin{tabular}{lcccc}
\hline\hline
 Source &    Method  & $S_{\rm 144~MHz}$ & $S_{\rm 1.425~GHz}$ & $\alpha$ \\
     &     &    mJy              & mJy               &          \\
\hline
AGN      &   & 181$\pm$18  &   15.1$\pm$0.8         &      \\  
Mini Halo & 1: Image subtraction & 92$\pm$9  &  8.9$\pm$0.4    & -1.02$\pm$0.05  \\
Mini Halo & 2: Uv subtraction &  90$\pm$9  & 8.7$\pm$0.4    &  -1.02$\pm$0.05 \\

Mini Halo  & 3: Double component fit  &  82$\pm$23& 11 $\pm$ 2&     -0.9$\pm$0.1\\
Mini Halo  & 4: HALO\_FDCA &    79$\pm$8   & 7.8$\pm$ 0.4 &  -1.01 $\pm$0.05 \\

\hline\hline

\multicolumn{5}{l}{\scriptsize Col 1: Source; Col 2,3: Flux density at 144 MHz and 1.425 GHz; Col 4: spectral index computed considering }\\

\multicolumn{5}{l}{\scriptsize  the emission detected above 2 $\sigma_{rms}$  in both images}

\end{tabular}
\label{tab:sources}
\end{table*}

\begin{figure}
\centering
\includegraphics[width=0.5\textwidth]{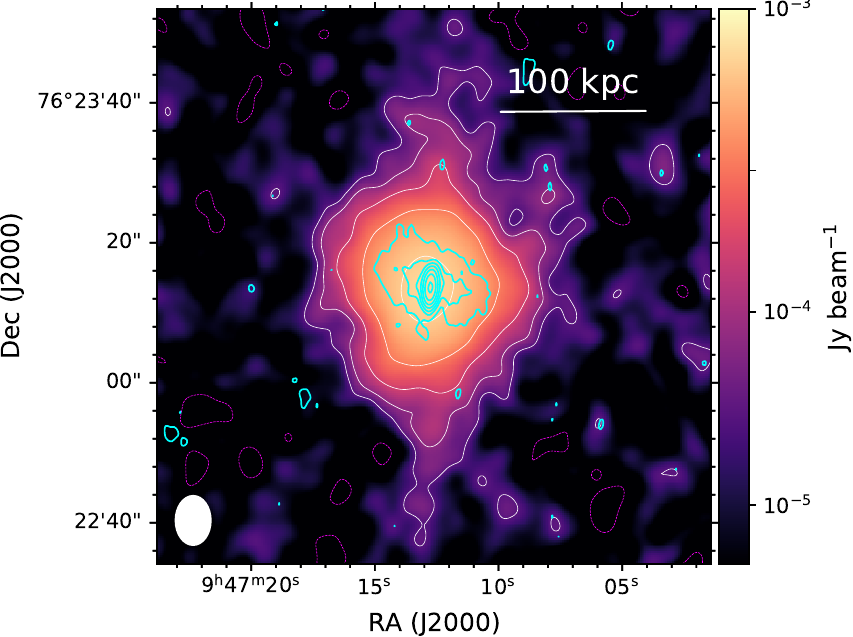}
\includegraphics[width=0.5\textwidth]{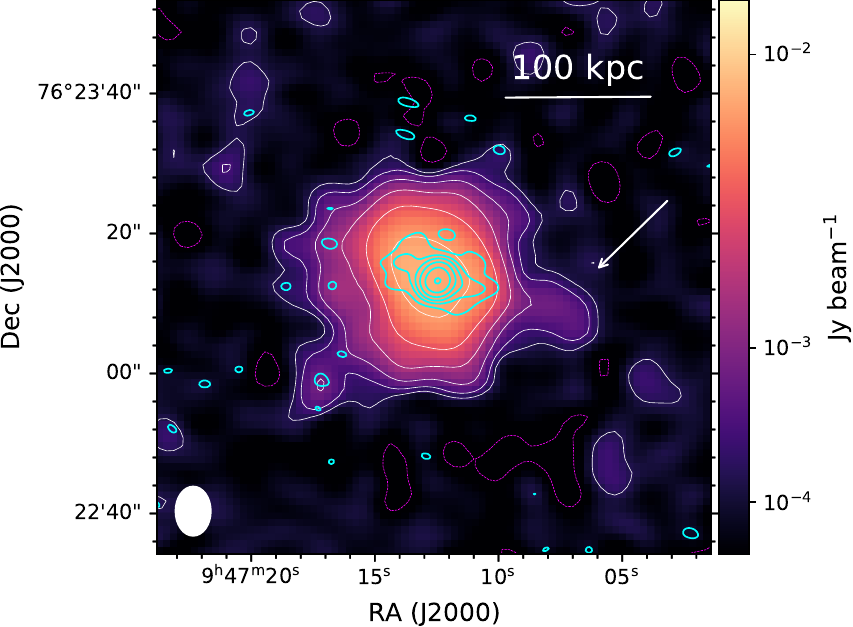}
\caption{Cluster RBS~797 at LOFAR and VLA frequencies. Top panel: VLA image at 1.425 GHz of the cluster RBS~797. Colours, white, and magenta contours refer to the image at resolution of 7\arcsec $\times$ 5\arcsec obtained after the subtraction of the central sources. The noise is $\sigma_{rms}= 0.02~\rm{mJy~beam^{-1}}$. White contours start at 2$\sigma_{rms}$ and increase by a factor 2 each. The -3$\sigma_{rms}$ contour is displayed in magenta. Cyan contours refer to the image of the central source obtained imposing an inner UV-cut of $10^4\lambda$ and start at 3$\sigma_{rms}=0.02~\rm{mJy~beam^{-1}}$, where the beam is 2.4\arcsec $\times$ 1.2\arcsec. Cyan contours increase by a factor 2.
Bottom panel: LOFAR HBA image at 144 MHz of the cluster RBS~797. Colours, white, and magenta  contours refer to the image at resolution of 7\arcsec $\times$ 5\arcsec obtained after the subtraction of the central sources. The noise is $\sigma_{rms}= 0.1~\rm{mJy~beam^{-1}}$. White contours start at 2$\sigma_{rms}$ and increase by a factor 2 each. The -3$\sigma_{rms}$ contour is displayed in magenta. Cyan contours refer to the image of the central source, obtained imposing an inner UV-cut of $8 \cdot 10^3\lambda$ and start at 3$\sigma_{rms}=0.3~\rm{mJy~beam^{-1}}$, where the beam is 3.1\arcsec $\times$ 2.1\arcsec. Cyan contours increase by a factor 2. The white arrow indicates the new western plume discovered with LOFAR.}
\label{fig:VLA_LOFAR_HL}
\end{figure}

\section{The radio emission in RBS~797}
\label{sec:res}

The total cluster emission at the central frequency of 144 MHz and at the resolution of 7.0\arcsec $\times$4.7 \arcsec is shown in Fig. \ref{fig:LOFAR_all}, obtained with the imaging parameters listed in Tab. \ref{tab:images}. The emission from the central AGN blends with the extended emission on cluster-core scale.
We confirm the presence of diffuse emission from the cluster central regions,
slightly asymmetric and elongated in the east-west direction. \\
The largest angular scale in the east-west direction is $\sim$ 50\arcsec, corresponding to 250 kpc
at the cluster redshift. In the north-south direction, the emission extends for $\sim$ 30\arcsec, corresponding to 150 kpc. 
The morphology  of the  mini halo detected by LOFAR at 144 MHz is slightly different from the image at 1.425 GHz published by \cite{Doria2012}, that is elongated in the north-south direction. Such morphology at 1.425 GHz is even more evident after the subtraction of the central AGN that we have performed in this work on the same data as those published by \cite{Doria2012} (see Fig. \ref{fig:VLA_LOFAR_HL}, upper panel).\\

LOFAR observations are sensitive to steep spectrum extended emission, and they have allowed us to discover megaparsec-scale emission around mini halos \citep{Savini19,Biava21b,Riseley22}. In RBS~797, instead, no additional emission on cluster-scale is detected. We refer the reader to \citet{Biava_sub}, where the upper limit to the cluster-scale emission is discussed.\\
At the cluster centre, the AGN emission is visible as a bright and slightly resolved component, but the resolution of 7.0\arcsec $\times$ 4.7\arcsec (corresponding to 35 $\times$ 23 kpc) does not permit one to disentangle the AGN emission from the mini halo emission.
Hence, we have reimaged the observations imposing an inner UV-cut of $8 \times 10^3 \lambda$, to filter out the emission on scales larger than $\sim$130 kpc. The image has a resolution of 3.1\arcsec $\times$ 2.1\arcsec, and is shown as cyan contours in the lower panel of Fig. \ref{fig:VLA_LOFAR_HL}. At this resolution, the AGN emission is resolved into a bright nucleus and two radio lobes, extending for 85 kpc in the east-west direction, i.e. the same directions as the E-W cavity pair detected in the X-rays.
The orientation of the lobes at 144 MHz is the same as in the 1.425 GHz image \citep{Gitti2006}. Similarly, the lobe extension at 144 MHz is confined within the inner shock $S_{in}$ detected by \cite{Ubertosi22} (see Fig. \ref{fig:LOFAR_all}).\\

We do not detect any further older lobes corresponding to the middle shock fronts $S_3-S_4$ and to  the outer shock fronts $S_5-S_6$ found by \citet{Ubertosi22}, which are located at an average distance of 80 and 130 kpc from the cluster centre (see Fig. \ref{fig:LOFAR_all}). However, we note an elongation of the mini halo towards the south-west, co-linear with the lobes of the AGN and extending up to 24\arcsec ($\sim$ 120 kpc) from the cluster centre. We refer to this component as a `plume' and discuss it further below.
The plume is labelled with a white arrow in Fig. \ref{fig:LOFAR_all} and \ref{fig:VLA_LOFAR_HL}, bottom panel.
We also note an extension of the mini halo towards the south-east, and ending at the position of the shock S5. However, a point-like source is visible at 1.425 GHz in that position \citet[see also][]{Biava_sub}, where an optical counterpart is found at the same position, hence we are not considering this extension as part of the mini halo.\\
From the high-resolution image, we have computed the flux density of the AGN within 3$\sigma$. This value is listed in  Tab. \ref{tab:sources}. In the same Tab., we also list the flux density of the mini halo.

\section{Disentangling the mini halo emission from the AGN}
\label{sec:fits}
Mini halos are often difficult to characterise, because of the  contamination from the central AGN. RBS~797 is likely one of the most complicated cases, as the central AGN is bright and resolved. In addition, it is known to have undergone multiple outbursts in different directions \citep{Gitti2006,Doria2012}, either coeval or in different epochs, that may further complicate the subtraction of the AGN emission. In order to study the emission of the mini halo, the AGN contribution must be removed. We proceeded in four different ways, that we detail below.

\subsection{Method 1: AGN subtraction}
We subtracted the central AGN and radio lobes flux densities, measured from the high-resolution image, above 3$\sigma_{rms}$
to the total flux density computed inside the $2\sigma_{rms}$ contours, computed from the low-resolution image. 
The mini halo integrated density at LOFAR frequencies corresponds to $S^{144~MHz}_{MH,1}=92\pm 9$ mJy, while in the VLA image we measure $S^{1.4~GHz}_{MH,1}=8.9\pm 0.4$ mJy,  which is within the range of values reported by \cite{Doria2012}. 
This is the simplest  method to account for the AGN emission, where neither the mini halo emission superimposed onto the AGN, nor the possible contamination of the AGN to the diffuse emission observed at low resolution are considered. We have used this method here to have a direct comparison with previous works.

\subsection{Method 2: AGN subtraction in the UV domain} 
The emission from the AGN can be in principle removed by subtracting the 
 corresponding visibilities in the UV plane. This has never been attempted in RBS~797, because of the difficulty in modelling the AGN emission. We investigate this method here, and compare the results with the other methods we have used.\\
 Specifically, both for VLA and LOFAR observations, 
 we have produced a high-resolution image of the central AGN by excluding the visibilities shorter than $10^4 ~\lambda$, corresponding to 100 kpc, to filter out the diffuse emission from the mini halo. We have used a uniform weighting scheme to further suppress the extended emission. Details on the imaging parameters are listed in Tab. \ref{tab:images} and the images are shown as cyan iso-brightness contours in Fig. \ref{fig:VLA_LOFAR_HL}.\\
 The models obtained from these images have been anti-Fourier transformed back in the visibility space and subtracted from the original data.
 UV-subtracted data have been reimaged at lower resolution ( $7 \arcsec \times 5 \arcsec $) using a Briggs weighting scheme \citep{Briggs} with robust=-0.5 (see Tab. \ref{tab:images} for further details).  The UV-subtracted images are shown in Fig. \ref{fig:VLA_LOFAR_HL}  in colours and white iso-brightness contours. We have used these images and measured the flux density within the 2$\sigma_{rms}$ contours.\\
 The resulting flux density is $S^{144~MHz}_{MH,2}=90\pm 9$ mJy at 144 MHz, and  $S^{1.4~GHz}_{MH,2}=8.7\pm 0.4$ at 1.425 GHz. These values are in overall agreement with the values  $S_{MH,1}$ derived above. \\
 This method allows us to better investigate the mini halo emission, though we cannot exclude either some residual contamination from the AGN or an over-subtraction of the AGN emission.
 The mini halo emission at LOFAR frequencies shows a spherical shape, though we notice a plume of emission towards the south-west, aligned with the inner lobes of the AGN.

Interestingly, the shock front $S6$ is located just outside the ``plume'' (see Fig. \ref{fig:LOFAR_all}).
 As the central AGN activity in this cluster is 
quite complex and multiple AGN outbursts have been found from X-ray and radio analysis \citep{Gitti2006,Ubertosi21}, we speculate that the plume detected at 144 MHz could be the remnant of an older ($\sim84$~Myr, \citealt{Ubertosi22} outburst from the central AGN. 

We note that the mini halo is elongated in the north-south direction at 1.4 GHz. \cite{Doria2012} already suggested a possible elongation in the north-south direction of the mini halo from the VLA image, but here it appears more evident, thanks to the subtraction of the central AGN from the UV data.
We also note that the ``plume'' towards the south-west detected at 144 MHz is not detected at 1.425 GHz. Considering a lower brightness limit of 2 $\sigma$ at 1.425 GHz and the mean surface brightness at 144 MHz, the plume emission should be characterised by a spectral index $\alpha < -1.4$.

\subsection{Method 3: Analytic profile: double fit}
We have tried a third method to estimate the flux density of the mini halo, following the approach proposed by \cite{Murgia09}. We have used the LOFAR image at 7.0\arcsec$\times$4.7\arcsec,  convolved the VLA image to the same resolution (see Tab. \ref{tab:images} for the LOFAR image), and fitted the total radio brightness profiles of the cluster with a central point source plus the radio mini-halo emission.  Then, we have computed the mean brightness per square arcsec within concentric annuli. We have considered all the annuli with a signal-to-noise ratio larger than or equal to 3 in the fit.\\
The radio emission of the cluster has been modelled as
\begin{equation}
    I(r)= I_{MH}(r) + I_{AGN}(r)
\label{eq:doubleFit}
\end{equation}
where $I_{MH}(r)= I_0 \exp^{-(r/r_e)}$ is the mini halo brightness profile,  and $I_{AGN}(r)= I_{AGN,0} \exp^{-(r^2/2\sigma_{AGN}^2)}$ is a Gaussian function used to describe the AGN brightness profile.
As explained in \cite{Murgia09}, the resolution of the observations and the sampling of the radial profile in annuli can affect the estimate of the best-fit parameters. Hence, we have taken into account these effects in the fitting procedure in the following way: we have produced 2-dimensional images with the same sampling in pixels as the observed images, changing the values of the free parameters $I_0,r_e,I_{AGN,0}, \sigma_{AGN}$. These images have been convolved with a Gaussian function having the major and minor axis equal to the observed beam major and minor axis, and a radial profile has been derived using the same annuli as for the observed images. 
The observed profiles, together with the best fit Gaussian and exponential profiles, are shown in Fig. \ref{fig:doubleFit} for the VLA (left panel) and LOFAR (right panel).
The values of the best fit of $I_0$ and $r_e,$ that we have obtained are deconvolved quantities and are listed in Table Tab. \ref{tab:expFit}. 
We note that, contrary to what has been obtained in the mini halos studied by \cite{Murgia2009}, the best-fit values of the mini halo change depending on the width of the annuli used for the radial profile, while the best-fit values of the Gaussian component are stable within the statistical errors of the fit.
This is likely due to the fact that the mini halo emission is detected at low signal-to-noise in the peripheral regions, and the outermost points are below or above the noise depending on the chosen averaging.
We have chosen to use the width of the annuli equivalent to half of the beam FWHM, i.e. 3.5 \arcsec, for a direct comparison with the mini halos studied by \cite{Murgia09}. The flux density of the mini halo, computed within 3$r_e$, is listed in Tab. \ref{tab:sources}, to have an immediate comparison with the results obtained from other methods. \\

\subsection{Method 4: Bayesian fit with HALO\_FDCA }
 \cite{Boxelaar_2021} have developed a Bayesian algorithm that allows one to fit circular, ellipsoidal, and skewed exponential surface brightness profiles to extended cluster radio sources.
 The package is called {\tt HALO FDCA} (Radio Halo Flux Density Calculator\protect\footnote{\url{https://github.com/JortBox/Halo-FDCA}}).
 The fit is performed on the two-dimensional image directly, using Markov chain Monte Carlo (MCMC) algorithm to explore the parameter space. We have investigated the circular and ellipsoidal profiles for LOFAR and VLA UV-subtracted images (Fig. \ref{fig:VLA_LOFAR_HL}, colours and white contours).
We refer to  \citet{Boxelaar_2021} for a detailed explanation and summarise here the relevant parameters only.
The surface brightness model is given by:
\begin{equation}
    I(r) = I_0 \exp^{-G(r)} ,
    \label{eq:exp}
\end{equation}
where $I_0$ is the central surface brightness and $G(r)$ a radial function. $G(r)= \left( \frac{|r^2|}{r_e^2}\right)^{0.5}$ for the circular model, while $G(r)= \left( \frac{x^2}{r_1^2} + \frac{y^2}{r_2^2}\right)^{0.5}$ for the elliptical models, where $r_e$ is the characteristic e-folding radius, and $r^2=x^2 + y^2$. 

We have applied the algorithm to the UV-subtracted images, at both 1.425 GHz and 144 MHz. The best-fit results are listed in Table \ref{tab:expFit}.
The flux density listed in the table refers to the emission within 3$r_e$, which is the 80\% of the total emission that would be derived from the analytic profile. This cut is introduced to account for the finite size of the sources, and is the one usually adopted in the literature.  We obtain $S^{144~MHz}_{MH,3}=79\pm 8$ mJy at 144 MHz, and  $S^{1.4~GHz}_{MH,3}=7.8\pm 0.4$ at 1.425 GHz, both for the circular and elliptical model.\\
The flux densities of the mini halo are also listed in Tab. \ref{tab:sources} for an easy comparison with the other methods we are using. We note that the flux density values obtained with this method are lower than those derived with the methods 1 and 2 described above. The reduced $\chi^2$ value ($\chi_r^2$) reported by the fit is 2.5 for the VLA image and 4 for the LOFAR image. \cite{Botteon22}  have analysed a sample of 309 clusters observed by LOFAR, and derived the fit to the radio halo in 89 of them using HALO\_FDCA, as we are doing in this work. The authors found 
a positive correlation between the signal-to-noise ratio (S/N) of the detection of the radio halo emission and the value of $\chi_r^2$ computed by the algorithm (See Fig. 7 of \citealt{Botteon22}).
The authors suggest that this trend is due to the presence of sub-structures that become significant at  high S/N, and that
cause a larger deviation from a smooth analytic profile.
In order to understand whether we could be subject to the same effect here, we have compared the S/N given by the algorithm for RBS797 to the S/N of the halos analysed by \citet{Botteon22}.

The mini halo of RBS797 is detected with
$S/N=73$ (VLA) and $S/N=94$ (LOFAR), and from \cite{Botteon22}, Fig. 7, clusters at this S/N have a $\chi_r^2>3$ and $\chi_r^2>5$, respectively. Hence, the high $\chi_r^2$ values derived for RBS~797 are consistent with those of halos detected at similar S/N. \\

\subsection{Comparison between different methods}

\begin{figure*}
\centering
\includegraphics[width=0.44\textwidth]{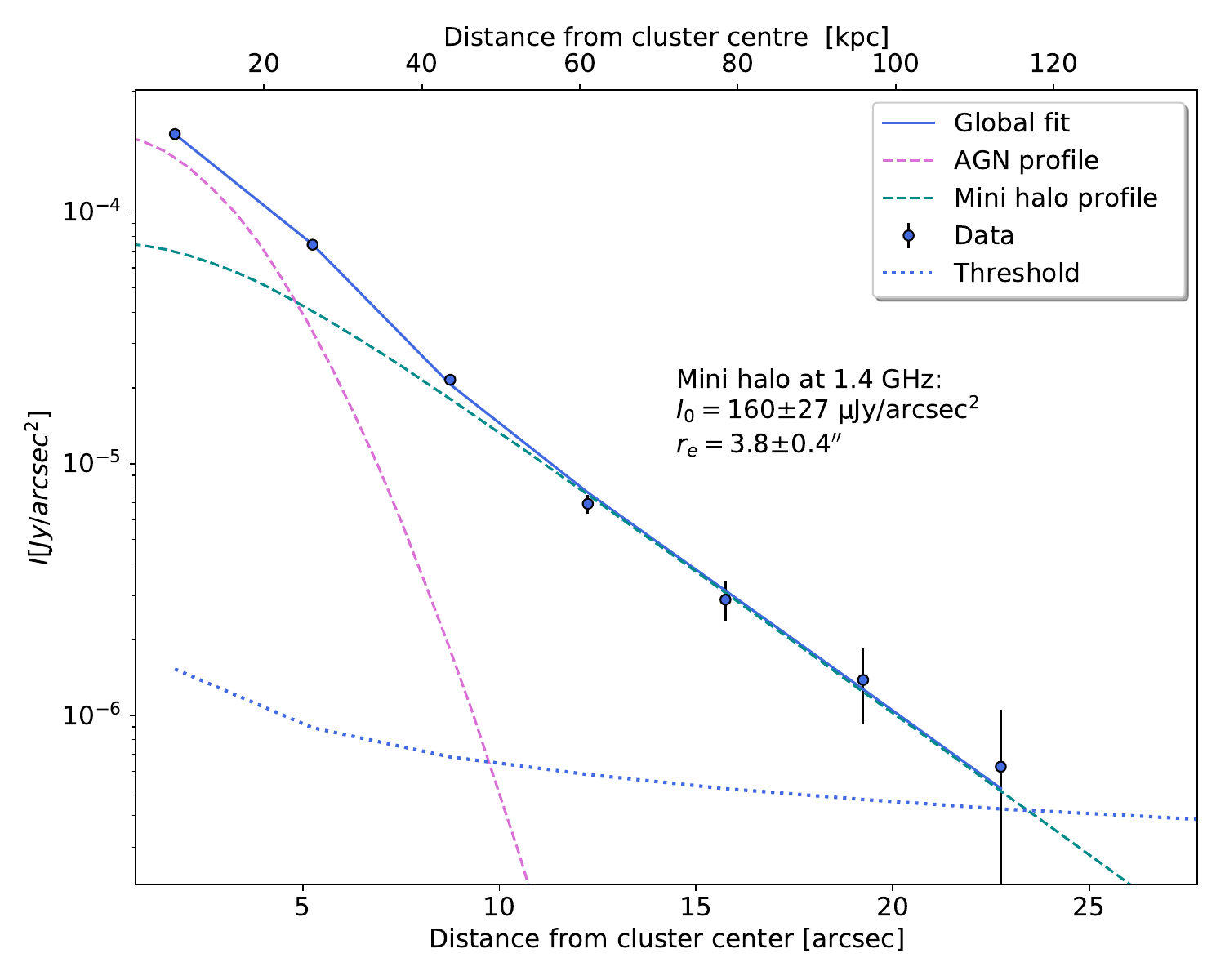}
\includegraphics[width=0.44\textwidth]{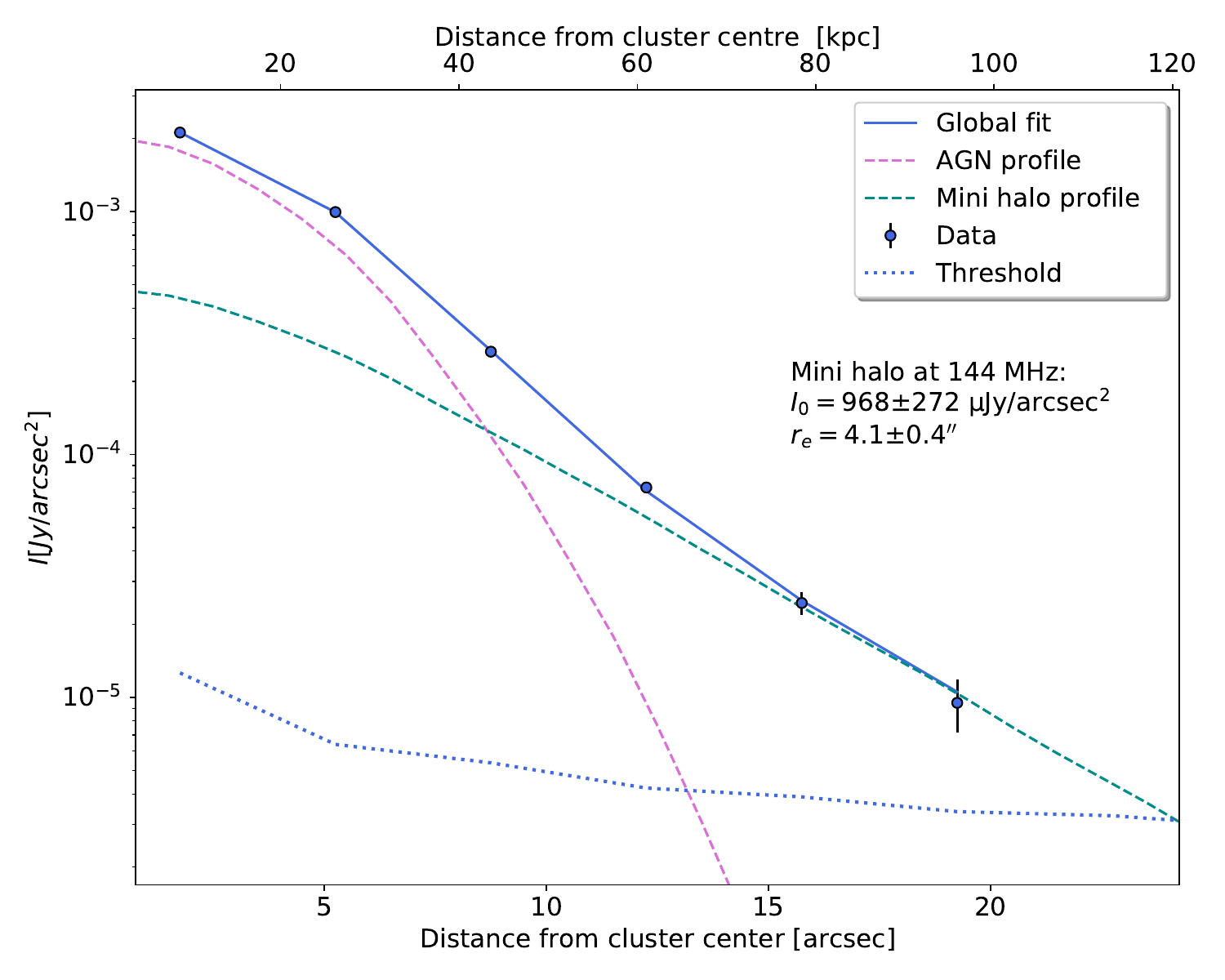}
\caption{Exponential plus Gaussian fit of the emission from RBS~797 (Eq. \ref{eq:doubleFit}) for VLA (left) and LOFAR (right) images. 
The blue solid line refers to the total best fit (mini halo plus AGN emission). Points refer to the mean brightness per square arcsec observed in each annulus, and the dotted blue line shows the thresholds we have considered (3$\sigma_{rms} / \sqrt{N_{beams}}$, with $N_{Beams}$ being the number of beams sampled in each annulus). The green line refers to the exponential mini halo fitted profile, the purple line to the gaussian AGN fitted profile. Best-fit parameters are listed in Table \ref{tab:expFit}.}
\label{fig:doubleFit}
\end{figure*}

The methods used to separate the AGN from the mini halo give a range of results, that reflect the difficulty in disentangling the two emissions.
Each of the methods we have used relies on some assumptions. Leaving method 1 aside, methods 2 and 4 start from UV-subtracted images, and rely on the assumption that the emission from the central AGN is not significantly affecting the residual mini halo emission, while method 3 starts from non-UV subtracted images, but it relies on the assumption that both the mini halo and the central AGN have a radial symmetry, from which they both deviate to some extent. Method 3, i.e. a fit with a Gaussian and exponential function, gives systematically higher values of $I_0$ than the other methods, and larger uncertainties on the flux density of the mini halo.
The reduced $\chi^2$ values computed by HALO\_FDCA are computed on a pixel basis, while the reduced $\chi^2$ values that we have listed for the double exponential plus Gaussian fit are computed on radial averages of the mini halo flux densities.
Hence, to compare the different fits, we have computed a second  reduced $\chi^2$ in the following way: we have taken the best-fit parameters found by HALO\_FDCA and we have computed 2D images of the mini halo at 144 MHz and 1.425 GHz, convolving the images with a Gaussian function having major and minor axis equal to the restoring beams. Then, we have divided the synthetic and observed mini halo in circular concentric annuli having width equal to half of the FWHM of the observing beam, and we have computed the reduced $\chi^2$ between these quantities. Using this method, one should be less sensitive to local substructures, and obtain values for the reduced $\chi^2$ that can be compared with the exponential plus Gaussian model fit. Using this approach, we obtain $\chi_r^2 = 1.3 $ for the VLA image, hence comparable with the reduced $\chi_r^2$ obtained from the Double Fit model. For the LOFAR image, we find reduced $\chi_r^2=5.4$, which is comparable to the value found by HALO\_FDCA and significantly higher than $\chi_r^2=0.9$, i.e. the value obtained with the Double exponential plus Gaussian model fit  (see
Table \ref{tab:expFit}.
Hence, we can conclude that the Double exponential plus Gaussian fit method provides the best description of the mini halo emission, i.e. it is difficult to subtract all and exclusively the emission from the central AGN, while taking it into account with a double fit provides a better modelling of the emission.
In addition, the double fit of exponential plus Gaussian component allows us to have a first estimate of the radial distance from the cluster centre where the mini halo emission is dominant with respect to the central AGN. From Fig. \ref{fig:doubleFit}, we can conclude that at a 10\arcsec  distance from the cluster centre, the contribution of the total radio emission from the mini halo is larger than the contribution from the AGN in both LOFAR and VLA images. At 13\arcsec  distance, the contribution of the central component is below the noise level in both LOFAR and VLA images. This is also consistent with the X-ray analysis by \cite{Ubertosi22}, as the radio-filled inner cavities extend up to 10\arcsec  from the cluster centre. 
In order to  define a region for a robust analysis of the mini halo spectral index, we have excluded a region of 10\arcsec ~ radius from the UV-subtracted images, and compared the flux density of the mini halo outside this region in UV-subtracted and non UV-subtracted images.
We obtain the following values for the LOFAR images: 21.4$\pm$0.6 mJy (non UV-subtracted)  and 21.0$\pm$0.6 (UV subtracted), and for the VLA images we obtain 3.8$\pm$0.2 mJy (non UV subtracted) and 3.6$\pm$0.2 mJy (UV-subtracted)\footnote{We note that only  statistical errors have been included in these estimates.}.
We can conclude that outside the inner 10\arcsec circular region the methods we have used to disentangle the mini halo emission and the AGN emission give consistent results.
In the following section, we use this information to investigate the spectral index radial trend of the mini halo.

\begin{table*}
\caption{Exponential fit results}
\renewcommand\arraystretch{1.2}
\begin{tabular}{lccccccc}
\hline\hline
\multicolumn{8}{c}{VLA - 1425 MHz}\\
\hline 
Method &  Model &   $\chi_r^2$ & $I_0$ & $r_1$  & $r_2$&  S  & P \\
    &       &              & $\mu \rm{Jy}/\rm{arcsec}^2$ & kpc&kpc  & mJy & $\rm{ 10^{24} W \, Hz^{-1}}$ \\

HALO\_FDCA    &Circular    & 2.5 & 51.2 $\pm$ 0.9 & 27.3$\pm$ 0.4  & - & 7.8$\pm$ 0.1 & 3.30$\pm$0.04 \\
HALO\_FDCA & Elliptical & 2.5 & 51.2$\pm$ 0.9 & 27.3$\pm$ 0.4 & 27.2$\pm$ 0.4  &7.8$\pm$ 0.1 & 3.30$\pm$0.04  \\
Double Fit  &   -     &   1.2 & 160$\pm$27 & 19$\pm$2 & - &  11$\pm$2 & 4.7$\pm$0.8 \\
\hline
\multicolumn{8}{c}{LOFAR - 144 MHz}\\
\hline
HALO\_FDCA    &Circular    & 4.8 &  688$\pm$9 & 23.7$\pm$0.2   & - & 78.8$\pm$0.9 & 33.4$\pm$0.4 \\
HALO\_FDCA    &Elliptical  & 4.8  &   692$\pm$9 &  24.2$\pm$0.3  & 23.1$\pm$ 0.3 & 78.7$\pm$0.8 & 33.3$\pm$0.3 \\
Double Fit  &   -     &   0.9  & 970$\pm$270 & 20$\pm$2 & - &  82$\pm$23&  35$\pm$8 \\

\hline

\multicolumn{8}{l}{Col 1: Method used; Col 2: model used for method HALO\_FDCA; Col 3: reduced $\chi^2$; Col. 4: central mini halo brightness;}\\
\multicolumn{8}{l}{Col. 5 (6): e-folding radius (radii for the elliptical model); Col. 7: Mini halo flux density within 3 times the e-folding radius;}\\
\multicolumn{8}{l}{Col. 8: Mini halo power corresponding to the flux density in Col. 7. and assuming $\alpha=1$. Only statistical errors are listed.  }
\end{tabular}
\label{tab:expFit}
\end{table*}

\section{Mini halo multifrequency analysis}
\label{sec:spix}

\begin{figure}
\centering
\includegraphics[width=0.5\textwidth]{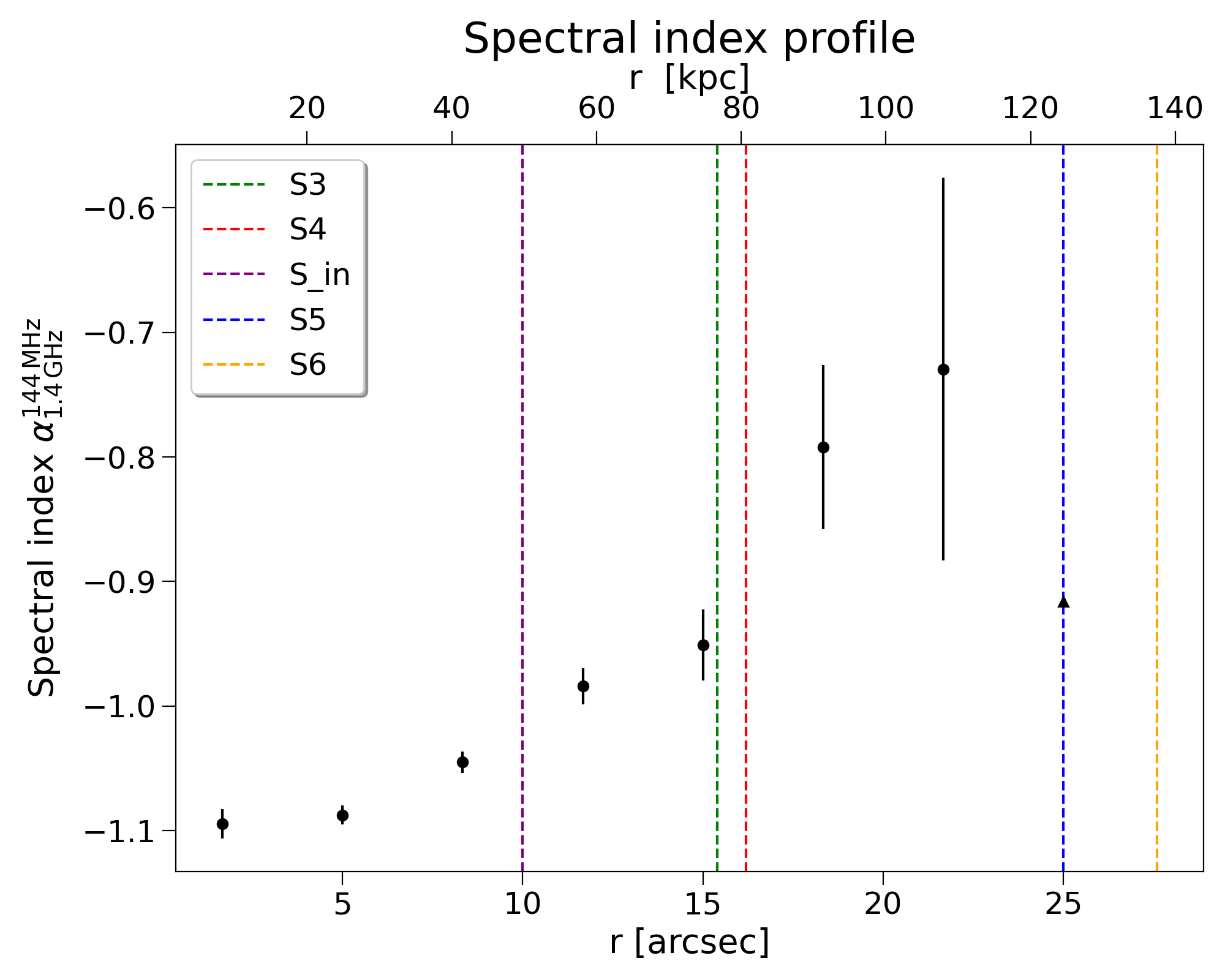}
\caption{Spectral index trend of the mini halo between 144 MHz and 1.425 GHz versus the distance from the cluster centre, assumed to be coincident with the AGN brightness peak in the LOFAR and VLA images, separately. Errorbars refer to statistical errors only. Vertical lines indicate the position of the shock fronts detected by \cite{Ubertosi22}.  }
\label{fig:spix_trend}
\end{figure}

The different methods we have used to separate the cluster diffuse emission from the AGN allow us to constrain the global spectral index of the mini halo.
The fluxes measured at 1.425 GHz and 144 MHz are listed in Table \ref{tab:sources}.
Despite the different flux densities obtained for the mini halo using different methods, the total spectral index values are consistent within the errors, and constrained to be in the range from $\alpha=-0.9\pm 0.1$ to $\alpha=-1.02 \pm 0.05$. 
This value is in agreement with the values reported in the literature for other mini halos \cite[see e.g.][]{Giacintucci2014,Riseley22}.\\
Using the UV-subtracted images, we can perform a resolved study of the mini halo spectral index, and investigate the connection between the radio emission and the models proposed for its origin.
In presence of a synchrotron break in the integrated radio spectrum, a radial steepening of the radio emission is expected by homogeneous re-acceleration models \citep{Brunetti01}. This has been observed in some radio halos \citep[e.g.][]{Bonafede22}, while it has never been detected in radio mini halos.\\ 
The multiple shock fronts detected  by \citet{Ubertosi22} could also re-accelerate the cosmic ray electrons and/or compress the radio emitting plasma and leave a clear imprint in the spectral index trend.\\
We have divided the mini halo emission in circular annuli having a width equal to half of the beam FWHM (i.e. 3\arcsec ) centred on the peak of the radio emission in the VLA and LOFAR image separately. Using the mean brightness in each annulus, we have computed the mean spectral index of the mini halo and the associated errors.
\\
In Fig. \ref{fig:spix_trend}, we plot the trend of the spectral index versus the distance from the centre of the mini halo at 1.425 GHz. Values for the spectral index are derived in all annuli where the average brightness at both frequencies is above the 2 $\sigma_{rms}$ threshold. 
Only statistical errors have been taken into account to visualise the spectral index trend. 
 Indeed, the systematic errors are due to the uncertainty of the flux scale that we have used, which would cause a shift of all the values towards either high or lower spectral index values. Since our aim here is to visualise the spectral index trend, the systematic errors have to be excluded.

The vertical lines indicate the position of the shock fronts as detected by \cite{Ubertosi22}.\\
The spectral index trend shows a progressive flattening from the cluster centre ($\alpha = -1.09 \pm 0.05 $) to the mini halo peripheral regions ($\alpha=-0.7\pm$ 0.2) out to a distance $r\sim 22$\arcsec (110 kpc) from the cluster centre.

\section{Discussion}
\label{sec:discussion}
The spectral index trend and the presence of shocks  are unusual features for mini halos. In the following, we discuss three possible scenarios to explain the observed properties of RBS~797.

\subsection{Re-acceleration by shocks}
The spectral index trend, steepening from the shock fronts downstream, is similar to the one observed in radio relics \citep[e.g.][]{Rajpurohit20}. In the case of relics, the steepening of the spectral index towards the cluster centre has been interpreted as particle ageing downstream after the re-acceleration by the shock at the relic outer edge. 

All the shocks detected in RBS~797 have a low Mach number $\mathcal{M}
\sim 1.2$ \citep{Ubertosi22}. Such low Mach numbers are typical of AGN-driven shocks in the ICM \citep[e.g.][]{Nulsen2005,Siemiginowska12}, and they are known to be
inefficient in accelerating particles from the thermal pool \citep[e.g.][]{Kang12,Vazza13,Botteon20}. However, 
if a population of pre-existing cosmic ray electrons is present, 
the problem of inefficient injection at weak shocks can be alleviated \citep[e.g][]{Kang12}.
In the case of RBS~797, a pre-existing population of CRe could be easily deposited in the ICM by the central AGN \citep[e.g.][]{Ignesti20}. \\
Making assumptions about the shock geometry and the ICM conditions at the cluster centre, we can check whether the size of the mini halo and its 
radial profile are compatible with particle re-acceleration by three shock waves.
We follow \cite{Kang12}, who assume plane-parallel shocks and stationary conditions, and derive the width $L_{rad}$ of the radio emitting region behind the shock, in the presence of a fossil population of cosmic ray electrons that are re-accelerated by the shock and that radiate their energy via synchrotron and Inverse Compton losses. From \cite{Kang12}, at the cluster's redshift, we have that:
\begin{equation}
L_{rad} \approx 765 \cdot \left( \frac{v_d}{10^3 \, {\rm km/s}}\right) \cdot \left( \frac{B_d^{1/2}}{B_{eff,2}^2} \right) \cdot \left( \frac{\nu}{\rm GHz} \right)^{-1/2}  \, {\rm kpc}
\end{equation}
where $B_d$ and $v_d$ are the downstream magnetic field and gas velocity, respectively.
The downstream magnetic field at distance $r$ from the cluster centre has been computed as $B_d= \sqrt{\frac{ 2*(C*B(r))^2 + B(r)^2}{3}}$, where $C$ is the shock compression factor, i.e. considering that only the magnetic field components parallel to the shock front are compressed.
$B_{eff,2}$ is the effective downstream magnetic field, that includes the equivalent strength of the cosmic background radiation with $B_{CBR} = 3.24(1+z)^2 \, \mu$G, so  $B_{eff,d}= B_d^2 + B_{CBR}^2$.
We have computed $L_{rad}$ for the inner, middle, and outer shocks using the 
average Mach numbers for each shock pair (inner, middle, and outer) to compute the downstream velocity and magnetic field values. For the latter, we have assumed a radial ``frozen-in'' scaling, i.e. $$B(r) = B_0 \left( \frac{n(r)}{n_0} \right) ^{2/3}.$$  
 The values of $n_0$ and $n(r)$ at the positions of the shocks have been taken by \citealt{Ubertosi22}.
Here, we have assumed a  central magnetic field $B_0 = 10 \, {\rm \mu G}$, in line with results published by e.g. \cite{Taylor02,Govoni06,Bonafede11,Vacca12}.
We note that $B_0 \sim 10-20 \, \mu$G are also expected in pure hadronic scenarios given the super-linear correlation found between the radio and X-ray surface brightness \citep{Ignesti20}.
 The average Mach number $\langle \mathcal{M} \rangle$, the downstream magnetic fields and velocities, and the resulting values for $L_{rad}$ for the three shocks at both VLA and LOFAR frequencies  are listed in Table \ref{tab:Lrad}.
\begin{table*}[]
    \centering
     \caption{Radio emission from the shocks}
     \renewcommand\arraystretch{1.2}
    \begin{tabular}{ c c c c c c c}
    \hline
    \hline
 Shock &  $\langle \mathcal{M} \rangle$ & $B_{d}$  &   $v_{shock}$  &  $v_d$  & $L_{rad}$ (VLA) & $L_{rad}$ (LOFAR)\\
  &                 & $\mu$G   &  km/s   & km/s &  kpc      & kpc \\ 
   
   \hline
    Inner  &  1.20$\pm$0.03   &  12$\pm$3  & 1540$\pm$150   & 1189$\pm$419 & 15$\pm$9& 46$\pm$29 \\
    Middle & 1.20$\pm$0.03    &  8$\pm$2   & 1498$\pm$150 &   1155$\pm$406&  22$\pm$12 & 69$\pm$37 \\
    Outer & 1.22 $\pm$0.03    &  4$\pm$ 1 &  1554$\pm$ 160   &  1172$\pm$402 &  29$\pm$11 & 92$\pm$34 \\
  \hline
  \hline
  \multicolumn{6}{l}{\scriptsize Col. 1 Shock label, Col. 2: average Mach number \citep{Ubertosi22}; }\\
  \multicolumn{6}{l}{\scriptsize Col 3: downstream magnetic field, assuming a central magnetic field of 10 $\mu$G;}\\
  \multicolumn{6}{l}{\scriptsize Col 4: downstream shock velocity; Col 5 and 6: radio emitting region behind the shocks}\\
   \multicolumn{6}{l}{\scriptsize  at VLA and LOFAR frequencies.}  \\
    \end{tabular}
   
    \label{tab:Lrad}
    
\end{table*}

The values of $L_{rad}$ that we have obtained are comparable, though slightly smaller at 1.425 GHz, to the width of the radio emission in between two consecutive shock fronts.
Although the numbers we have computed depend on the assumed $B_0$,  higher values of $B_0$ would give lower values of $L_{rad}$, while the maximum values of $L_{rad}$ are $L_{rad}\sim 30$ kpc for a constant magnetic field with a strength of 3 $\mu$G. Hence,  the values of $L_{rad}$ would not change significantly for other values of $B_0$ .We note that $L_{rad}$ would increase once the curved geometry of the shock and projection effects are considered, alleviating the  discrepancy.\\

 \subsubsection{Radio mini halo emission by shock re-acceleration}
In this section, we compute the radio profile that would originate by the three shocks - inner, middle and outer - assuming a pre-existing population of seed electrons.

Assuming that the re-accelerated electrons do not affect the shock dynamics, the downstream electron spectrum ($N_d(\gamma)$) will depend  on the energy distribution 
and spatial distribution of the seed electrons.
 Following \citet{Blandford87}, we have:
\begin{equation}
\label{eq:reacc}
 N_d(\gamma)= (\delta +2) \gamma^{- \delta} \int_{\gamma_{min}}^{\gamma} N_u (\gamma) \gamma^{\delta-1} d \gamma
\end{equation}
where $\delta$ is the flattest between slope of the spectrum of the seed electrons ($N_u(\gamma)\propto \gamma ^{- \delta_s}$) and the slope of the spectrum produced by direct shock acceleration, which depends on the shock compression factor $C$ according to $\delta_c= \frac{C+2}{C-1}$.
$\gamma_{min}$ is the maximum between the minimum energy of the seed electrons and the minimum energy of particles that can be accelerated by the shocks, and $N_u(\gamma)$ is the upstream (seed) electron energy spectrum (we refer the reader to \citealt{Markevitch05} for further details).\\
Using eq. \ref{eq:reacc}, the energy density of the re-accelerated electrons is written as follows:
\begin{equation}
\epsilon_d=\int_{\gamma_{min}}^{\gamma_{max}} N_d(x) m_e c^2 x dx.
\end{equation}
We note that $\epsilon_d$ depends on the energy spectrum of the seed electrons, which is unknown. However, the spectral shape of the re-accelerated (downstream) electrons, at energy larger than the maximum energy of the seed (up-stream) electrons will be a power-law with a slope that only depends on the shock Mach number. In this case, the shape of the seed (up-stream) electron spectrum would only affect the normalisation of $N_d$, i.e. of the re-accelerated electrons, and therefore we have:
$$\epsilon_d(\gamma)\propto \epsilon_u(\gamma).$$
In the following, we do not investigate the details of the spectrum of the seed electrons, and assume a spatial distribution of the seed electrons that follow the thermal gas distribution:
$$
\epsilon_d(r) \propto   n(r).
$$
Under these assumptions, we can only attempt to reproduce the radial shape of the observed radio emission, minus a normalisation factor.
 
We note that since the shocks in RBS~797 have a low Mach number, the re-accelerated electrons will produce a very steep synchrotron spectrum, with $\alpha= \frac{\delta-1}{2}\sim 5$. However, because of the magnetic field compression, the frequency at which the electrons will emit the most is given by $$\nu_{p} \propto \gamma^2  B \sim C^{4/3},$$  (see e.g. \citealt{Markevitch05} and ref. therein), which would produce a shift of a factor $\sim 1.4$ in the observed synchrotron spectrum\footnote{If the seed electrons are modelled as a relativistic gas, the compression factor increases  to $C=1.6$  and the shift in $\nu_{p}$ would be $\sim$ 1.9 }. 

Considering all the assumptions above, we could be in the lucky situation where we are observing at LOFAR and VLA frequency the spectrum of the seed electrons, shifted to higher energies. Hence, we proceed computing the mini halo radial profile that would result from the superposition of the three shocks. \\
 
For each shock, we have considered that the re-accelerated electrons - having  energy density $\epsilon_d$ - emit over a distance equal to $L_{rad}$ (see Tab. \ref{tab:Lrad}).
We focus on the $L_{rad}$ at 1.4 GHz, to check whether the re-acceleration from three shocks can explain the radio decrement from the centre to the periphery of the mini halo.
As $L_{rad}$ decreases at higher frequencies, reproducing the profile at the highest observing frequency is the main challenge of the model.

To derive the surface brightness radial profile under  stationary conditions,  we have integrated $\epsilon_d$ along the line of sight.
In order to compare the profile to real observations,  we have smoothed the analytical profile with a Gaussian function having the FWHM corresponding to the FWHM of the restoring beam of the VLA UV-subtracted image (see Tab.\ref{tab:images}). 
 Finally, we have normalised the smoothed analytic profile of the radio brightness to the observed 
best-fit value at 1.4 GHz (see Sec. \ref{sec:fits}) at the distance of 130 kpc from the cluster centre, i.e. at the distance of the outer shock.
This normalisation is needed to account for the unknown energy distribution of the seed electrons, as explained above.

\begin{figure}
\includegraphics[width=\columnwidth]{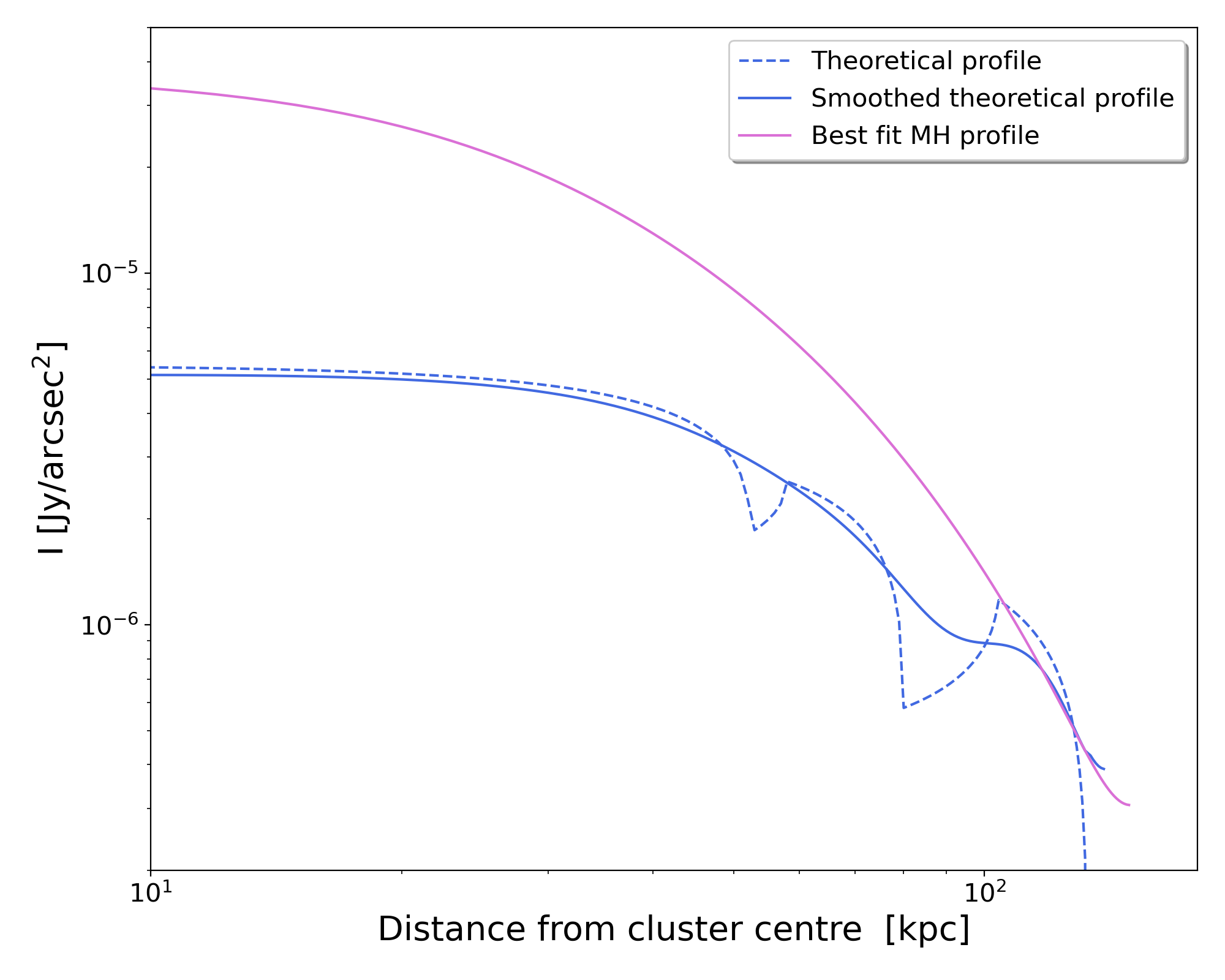}
\caption{Best fit mini halo profile and theoretical profile computed considering
the re-acceleration from the three shock waves. Blue dotted line is the theoretical shock profile, blue continuous line is the theoretical shock profile after smoothing to mimic the effect of beam convolution, rose line is the best-fit profile obtained with HALO\_FDCA  on VLA data at 1.425 GHz.}
    \label{fig:shock_profile}
\end{figure}

In Fig. \ref{fig:shock_profile}, we show the best-fit mini halo profile (obtained from the method HALO\_ FDCA, see Tab. \ref{tab:expFit}) and the theoretical profile that would result from shock re-acceleration at 1.4 GHz. 
It is clear that the re-acceleration by shocks alone cannot account for the observed radio emission. In particular, at a distance of 10 kpc from the cluster centre, the two profiles differ by a factor of 6. \\
Hence, to account for the radio profile of the mini halo, one should have a radial distribution of the seed electrons that is much more peaked towards the centre than the distribution of thermal electrons. 
For instance, in a scenario where the seed particles are generated by hadronic interactions between cosmic ray protons and thermal protons, the profile of the seed electrons would decline not as $n_e(r)$ but as $N_u(r)\propto \frac{n_e(r)}{r}$ as a consequence of the secondary CRe diffusion throughout the cluster \citep[see e.g.][and ref. therein]{Ignesti20}. This could reduce the factor 6 difference between the observed and analytic emissivity profiles at the cluster centre.

A detailed modelling of the energy and spatial distribution of the seed electrons is beyond the scope of this work.\  However, it is unlikely that the effect of shock re-acceleration alone would be able to reproduce the observed profile.

\subsubsection{Polarisation from shocks}
From the analysis performed above, we can conclude that it is unlikely that shock re-acceleration with a constant efficiency 
would  be able to explain the radial radio profile of the mini halo.
However,  the spectral features observed in  RBS797 
could be explained assuming that shocks are propagating onto a pre-existing radio emitting mini halo. Though the effect of the shocks could be more or less important, depending on the energy and density of the mini halo CRe, shocks will compress the gas and boost the magnetic field components parallel to the shock fronts by a factor $C =1.3$ for $\langle \mathcal M \rangle = 1.2$. This compression will increase the polarisation of the radio plasma.
Following \cite{Ensslin98}, in the case of weak magnetic field, the expected polarisation fraction, $p_{0}$ should be $p_{0}\sim 23\%$, assuming that the shocks are seen  at a viewing angle of 90 degrees between the shock normal and the line of sight. This emission will undergo external depolarisation from the intra-cluster medium, resulting in an observed polarisation $P$ at the wavelength $\lambda$ (see e.g. \citealt{OSullivan12}):
$$
P=p_0 e^{-2 \sigma_{RM}^2 \lambda^4} e^{2i (\Psi_0 + RM \lambda^2)}.
$$
Here, $RM$ is the Faraday Rotation measure due to the cluster magnetic field,  $\sigma_{RM}$ is the dispersion in Faraday $RM$ observed over the source, and $\Psi_0$ is the intrinsic polarisation angle on the radiation. 
Assuming a $\sigma_{RM}\sim 50- 300 \; \rm{rad/m^2}$ (e.g. \citealt{Taylor02,Govoni2009,Vacca12}), we derive that at 6 GHz the observed polarisation should be $\sim 7-20 \%$ of the total intensity emission.
We note that \citet{Gitti2006} have reported the detection of polarised emission at 1.425 GHz, in a region extending up to 10\arcsec $\times$ 15\arcsec from the centre of the cluster. We have re-imaged the VLA data at 1.425 GHz in polarisation after the subtraction of the AGN, and the detection of polarised emission results just above the 2$\sigma$ noise level. Therefore, we conclude that future deep observations would be needed, to probe whether the shock waves have compressed the ICM magnetic field.

\subsection{Turbulent (re)acceleration in a high magnetic field}

A second scenario to explain the observed properties of the mini halo is that re-acceleration by turbulence is taking place in a cluster with a high magnetic field.

The spectral index trend is  the opposite of what has been found in giant radio halos \citep[e.g.][]{Bonafede22,Rajpurohit20} and expected from turbulent re-acceleration models, in the case that the magnetic field $B$ is $B< \frac{B_{CBR}}{3}$ \citep{Brunetti01}.
Indeed, the synchrotron cut-off frequency $\nu_c$ is:
\begin{equation}
\nu_c \propto \frac{B}{(B^2+ B_{CBR}^2)^2}
\end{equation}
which decreases for $B < \frac{B_{CBR}}{\sqrt{3}}$, causing a steepening of the synchrotron spectrum in a radial declining magnetic field.

 Although magnetic fields in cool core clusters are largely unconstrained, having a central magnetic field $B > \frac{B_{CBR}}{\sqrt{3}} \sim 3.43\, \mu$G at the redshift of RBS~797, is in agreement with current works \citep[e.g.][]{Taylor02, Vacca12, Bonafede11,Osinga22}.

Assuming a central magnetic field $B_0$ and a radial declining magnetic field $$ B(r) = B_0 \left( \frac{n_e}{n_0} \right)^{2/3} $$
one would expect a radial flattening  up to a distance $r^*$ where $B(r^*)= 3.43 \, \mu$G, and then a radial steepening of the synchrotron spectrum.

For $B_0=10 \, \mu$G, assuming the $\beta$-model parameters by \citep{2011cavagnolo}, one obtains $r^* \sim $ 15 \arcsec, which is at odd with the observed spectral index radial trend. In order to have $r^*>$ 22 \arcsec, one would need $B_0> 15 \,\mu$G, and  one would expect to see a steepening of the radio spectrum at larger distances. 
We note that this scenario would explain the radial spectral flattening independently of the presence of the shocks detected in the X-rays. Hence, one would expect to observe  similar behaviour in all mini halos hosted by clusters with similar magnetic field strengths. 
Possibly, deeper observations and studies on other mini halos could confirm or reject this scenario.

\subsection{Hadronic scenario with energy dependent spatial diffusion coefficient}

A third possibility to explain the radial spectral index trend is to assume a hadronic scenario, where CRe are produced as a consequence of thermal proton-CRp interactions \citep{PfrommerEnsslin04}, and are subject to a spatial diffusion coefficient $D(p)$ that is an increasing function of the CRp momentum $p$ \citep[see][sec. 3.2, where the diffusion of CRp in the ICM is discussed.]{BJ14}. 
In this scenario, the most energetic CRp would be able to diffuse faster than the lower energetic CRp, reaching larger distances from the injection site (e.g. the central AGN). As a consequence, one would have an increasing density of more energetic particles at a further distance from the AGN, and the CRp spectrum would show a radial flattening. As the CRe spectrum depends on the CRp spectrum, this would explain the observed radial flattening of the synchrotron emission.\\

Under simplified assumptions, we can constrain the 
 energy dependence and normalisation of the diffusion coefficient  $D(p)$.  \\ 
 The propagation of CRp in galaxy clusters is diffusive, and can be described by the transport equation (see \citealt{BlasiColafrancesco99} for details). 
 The solution of the transport equation at time $t< t_0$, being $t_0$ the time between the start of the injection and the observing time, is given by (see \citealt{BlasiColafrancesco99}):
\begin{equation}
n(p,r)_{|t < t_0}= \frac{Q(p)}{2 \sqrt \pi^3 D(p) r} \int_{r/r_{max}}^{\infty} e^{-y^2} dy
\label{eq:diff}
\end{equation}
where $n(p,r)$ is the number of CRp at distance $r$ from the source of injection (the AGN),  having momentum $p$, and $Q(p)$ is the injection spectrum.  The lower limit of the above integral is also a function of $p$, since the maximum injection scale, $r_{max}$, i.e. the maximum distance that a CRp with momentum $p$ can can diffuse away from the source in the time $t_0$, depends on $p$ according to:
\begin{equation}
r_{max}= \sqrt{4 D(p) t_0}.
\label{eq:rmax}
\end{equation}
\\
If we are in the limit $\frac{r}{r_{max}} \ll 1$, i.e. at distances much smaller that the injection scale, the integral above gives a stationary solution:
\begin{equation}
n(p,r)_{|t < t_0} = \frac{Q(p)}{4 \pi r D(p)}; \qquad r \ll r_{max}.
\end{equation}
{\bf Assuming that $D(p)= D_0 p^{k}, k>0$} and $Q(p) \propto p^{\delta}$,
we obtain a constant spectrum of the CRp and hence of the electrons responsible for the radio synchrotron emission:
\begin{equation}
n(p,r)_{|t < t_0} \propto \frac{p^{\delta -k}}{r};  \qquad r \ll r_{max}.
\end{equation}
From this, a constrain on $k$ can be put, assuming a typical AGN injection spectrum,  $\delta < -2$. Since we observe a constant spectrum $\alpha \simeq -1.1 $ up to 30 kpc from the cluster centre, we can constrain  $k \leq 0.2$.\\
At distances $r \geq r_{max}$, the spectrum of the CRp (hence of the radio-emitting electrons) will flatten, as the contribution of the integral in Eq. \ref{eq:diff} will increase with $p$.\\
Using eq. \ref{eq:rmax}, we can constrain the value of the diffusion coefficient $D(p)$ at a fixed energy that corresponds to the particles emitting at 144 MHz. Assuming a magnetic field of $B \sim 1 -10  \mu G$, and that through hadronic collisions 
protons with energy $E_p$ produce electrons with energy $E_e \sim E_p/10 \sim $ (see \citealt{BlasiColafrancesco99}),
the energy of the CRp that decay in radio-emitting CRe is of the order of $E_p \sim 50$ GeV. Hence, as we observe a flattening of the spectrum at distances larger than 30 kpc, we have:
$$
D(p)_{|E_p = 50 {\rm GeV}} < 10^{28}  \frac{r_{max}^2}{(30 {\rm kpc)^2}} \left( \frac{t_0}{500 {\rm Myr}} \right)^{-1} \; {\rm cm^2/s}.
$$
\\
A detailed modelling of this  process is beyond the aim of this work, and the constraints that we have obtained here on $D(p)$ depend on the assumed dependence of $D(p)$ from $k$ and on $t_0$. We also stress that we are assuming a constant magnetic field value, and that we are ignoring the possible effects that residual emission from the AGN could have at distances $r \sim 30 kpc$. 
A better sampling of the radio spectrum would be needed to derive more stringent constraints.

We note that  the flattening of the spectral index is not correlated to the presence of shocks in this scenario, and one would expect to  detect a similar spectral behaviours in other mini halos as well.

The three models we have outlined do not exhaust necessarily all the possibilities to explain the peculiar observational features of the mini halo 
in RBS~797. A test to  these models would be provided by obtaining constraints on the mini halo polarisation properties and on the cluster magnetic field.
Unless the magnetic field is larger than 15 $\mu$G at the cluster centre, 
the central AGN  has a role in shaping the spectral properties of the mini halo.\\

The idea that the properties of mini halos can be linked to the central AGN activity is not new.  
\citet{Fujita12} have investigated the radial profiles of  mini halos  assuming that the synchrotron emission originates from the CRp that heat the cluster cool core.

\citet{Bravi15} also have suggested a common origin of radio mini halos and gas heating by AGN-induced turbulence in cool cores. Later, \citet{Gendron17} have studied the emission of the mini halo in the Perseus cluster, finding that the mini halo seems to be influenced by both the AGN activity and the sloshing motion of the host cluster. The authors also found filamentary structure within the radio diffuse emission. Although filamentary structures are visible in AGN lobes when observed at high resolution and sensitivity \citep[e.g.][]{Maccagni20,Ramatsoku20,BrienzaNEST}, \citet{Gendron17} note that they resemble those observed in radio relics in merging galaxy clusters \citep[e.g.][]{Rajpurohit22}.\\
The connection between the AGN activity and the mini halo emission has been further investigated by \cite{Richard-Laferriere2020} who found a correlation between the radio power of the mini halo at 1.4 GHz and the X-ray cavity power\footnote{This correlation has been derived using only the inner X-ray cavities of systems with multiple cavities}. These authors conclude that AGN feedback could be a key-mechanism to understand the origin of mini halos in cool-core clusters. 
Indeed, the duty cycle of AGN in cluster centres is quite fast \citep[e.g.][]{Biava21a,Ubertosi21}, meaning that particles are injected in the medium by the AGN very often.
\cite{Richard-Laferriere2020} suggest that the main driver for the creation of mini halos could be
 the turbulence injected in the medium by the AGN, while sloshing motions could determine the overall shape of the mini halos.
 
Our multifrequency study on RBS~797 adds another piece to this picture, as the detection of multiple shock fronts connected with the AGN activity \citep{Ubertosi22}, the absence of core-sloshing evidence \citep{Biava_sub}, and the spectral index trend which is flattening towards the mini halo outskirts suggest that the properties of the mini halo depend on the AGN activity.

\section{Conclusions}
\label{sec:concl}
We have analysed the LOFAR HBA observations of the galaxy cluster RBS~797. 
RBS~797 is known to host a powerful AGN at its centre, and two equidistant and perpendicular cavity systems discovered with X-ray observations \citep{Schindler2001,Ubertosi21} that indicate either the presence of a binary AGN or a fast re-orientation of the jets (on timescales shorter than 10 Myr) that produced two different outbursts \citep{Ubertosi21}. Using deep {\it Chandra} observations, \citet{Ubertosi22} have discovered multiple shock fronts around the central AGN, connected with the AGN activity.\\
The cluster is known to host a mini halo \citep{Gitti2006}.\ However, little is known about its properties because of the difficulties in disentangling its emission from the AGN emission. \\
LOFAR observations have confirmed the presence of diffuse emission extending for $\sim$22\arcsec (110 kpc) from the cluster centre, while no diffuse emission is detected outside the core \citep{Biava_sub}.
We have studied the system using a LOFAR HBA observation with archival VLA observations and the recent X-ray analysis by \cite{Ubertosi22}. Our results can be summarised as follows:
\begin{itemize}
\item we have used different methods to separate the emission of the AGN from the mini halo emission, and we found that the mini halo is brighter than the AGN emission outside the inner 10\arcsec, both at 1.425 GHz and 144 MHz.
The mini halo at 144 MHz is not more extended than at 1.425 GHz down to the sensitivity level of our observations, and it is slightly elongated in the north-east south-west direction, opposite to  the north-south orientation found at 1.425 GHz.\\
\item We detect a plume of radio emission, co-linear with the western lobe detected at 144 MHz and extending up to the position of the shock S6. We derived a spectral index limit $\alpha < -1.4$ between 144 MHz and 1.425 GHz We speculate that the emission could be linked to the past AGN activity.
\item The mini halo emission can be fitted by an exponential profile. Once the central AGN emission was subtracted from the UV data, we determined that the best-fit parameters have a central brightness of $I_0=0.69 \pm 0.09 \; {\rm mJy/arcsec^2} \; [51 \pm 3 {\rm \mu Jy/arcsec^2}] $  and an e-folding radius $r_e=23.7 \pm 0.2\; [27.3 \pm 0.4]$ kpc at 144 MHz [1.425 GHz]. Its total power is $P_{144\, {\rm MHz}} = ( 3.3 \pm 0.3) \cdot 10^{25} \, {\rm W/Hz}$, and $P_{1.425\, {\rm GHz}} = ( 3.3 \pm 0.2 ) \cdot 10^{24} \, {\rm W/Hz} $;\\
We obtained different best-fit parameters if we fitted for a mini halo exponential profile plus a Gaussian profile to the non-subtracted images.
Specifically, we determined $I_0=0.97 \pm 0.27 \; {\rm mJy/arcsec^2} \; [160 \pm 27 {\rm \mu Jy/arcsec^2}] $  and an e-folding radius $r_e=20 \pm 2\; [19 \pm 2]$ kpc at 144 MHz [1.425 GHz]. The corresponding powers are $P_{144\, {\rm MHz}} = (35 \pm 8) \cdot 10^{25} \, {\rm W/Hz}$, and $P_{1.425\, {\rm GHz}} = ( 4.7 \pm 0.8 ) \cdot 10^{24} \, {\rm W/Hz} $.\\ 

\item We have analysed the spectral index profile of the mini halo between 144 MHz and 1.425 GHz, finding a progressive flattening towards the mini halo outer regions. The spectral index ranges from $\alpha=-1.09 \pm 0.05$ at the cluster centre to $\alpha=-0.7 \pm 0.2$ at 22\arcsec (110 kpc) from the cluster centre. A 2$\sigma$ lower limit was derived in the outermost annulus ($\alpha > -0.9$) at 25\arcsec (125 kpc) from the cluster centre. The derived spectral index trend and the presence of multiple shock fronts in the cluster point to a connection between the AGN activity and the radio properties of the mini halo. 

\item We have computed the width of the radio emission behind each shock detected in the X-rays, assuming a plane-parallel geometry and stationary conditions, and assuming a magnetic field profile with a central value of 10 $\mu$G and a  flux-freezing radial scaling. We have computed the projected surface brightness radio profile that would result from the re-acceleration by the three shock fronts,  assuming that the seed electrons have a radial distribution that follows the thermal gas distribution. We have  smoothed the derived profile with a Gaussian function to mimic the effect of the beam. 
We conclude that it is unlikely that the re-acceleration from shocks can  reproduce the radial brightness of the mini halo  alone, although  a detailed modelling of the emission considering the profile of the seed electrons should be performed. 

\item  We propose a first scenario where the AGN-driven shocks propagate onto an already existing mini halo, re-accelerating and compressing the radio plasma and causing the observed radial spectral flattening.
In this scenario, we expect that the polarisation from the shocks should be 8-20 \% of the total intensity emission at 6 GHz, which could be detected by future sensitive observations.

\item We also investigated the possibility that the radial flattening of the mini halo is caused by turbulent re-acceleration in the presence of a high magnetic field. We determined that the central magnetic field $B_0$ should be larger than 15 $\mu$G to account for the observed spectral flattening out to 22\arcsec from the cluster centre. In this case, we would also expect a steepening of the spectral index at the radius where the magnetic field becomes lower than $\sim 3.4 \;\mu$G.

\item A third possibility to explain the radial spectral flattening is to assume a pure hadronic scenario, where CRp are injected in the ICM by the AGN activity,  and a diffusion coefficient of CRp that depends on the CRp momentum. Using a simplified model, we have been able to put some constraints of the diffusion coefficient normalisation and on its energy dependence.
In this scenario, the shocks would not be affecting the properties of the mini halo, but the AGN activity would still have an essential role injecting the CRp in the medium.

\end{itemize}
Other models might be possible to explain the observed properties of the radio emission in RBS~797, and a detailed modelling of the proposed models is beyond the scope of this work. However, we  would be able to understand the role of the AGN and of the shocks with deep polarisation observations and obtain constraints on the magnetic field of the cluster.\\

\section*{Acknowledgments} 
The authors thank the referee, Yutaka Fujita, for his timely and detailed report.
AB, NB, and CJR acknowledge support from the ERC Stg n 714245 ``DRANOEL". AB and CS acknowledge support from the MUR grant FARE ``SMS".
AI acknowledges support from the European Research Council (ERC) under the European Union's Horizon 2020 research and innovation programme (grant agreement No. 833824).
MB acknowledges support from the agreement ASI-INAF n. 2017-14-H.O and from the PRIN MIUR 2017PH3WAT “Blackout”.

RJvW acknowledges support from the ERC Starting Grant ClusterWeb 804208.

NL is supported by the Dutch Black Hole Consortium (with project number 1292.19.202) of the research programme NWA which is (partly) financed by the Dutch Research Council (NWO). 
AI acknowledges support from the European Research Council (ERC) under the European Union's Horizon 2020 research and innovation programme (grant agreement No. 833824).

These data were (partly) processed by the LOFAR Two-Metre Sky Survey (LoTSS) team. This team made use of the LOFAR direction independent calibration pipeline (https://github.com/lofar-astron/prefactor) which was deployed by the LOFAR e-infragroup on the Dutch National Grid infrastructure with support of the SURF Co-operative through grants e-infra 160022 e-infra 160152 (Mechev et al. 2017, PoS(ISGC2017)002).

\bibliographystyle{aa} 
\bibliography{biblio}

\end{document}